\documentclass[preprint,aps,10pt]{revtex4-1}
\usepackage{amssymb,latexsym,amsmath,cancel,graphicx,epstopdf}
\usepackage{color}
\def\be {\begin{equation}}
\def\ee {\end{equation}}
\def\mn {{\mu\nu}}
\def\ba {\begin{eqnarray}}
\def\ea {\end{eqnarray}}
\def\nn {\nonumber}
\def\cm {{\cal M}}
\def\cl {{\cal L}}
\def\la {\langle}
\def\ra {\rangle}
\def\del {\partial}

\def\om {\omega}

\def\vq {\vec q}
\def\vk {\vec k}
\def\vp {\vec p}

\def\de {\delta}
\def\Gm {\Gamma}
\def\De {\Delta}
\def\Sg {\Sigma}
\def\Lm {\Lambda}
\def\sg {\sigma}

\begin{document}
\title{In-medium viscous coefficients of a hot hadronic gas mixture}

\author{Utsab Gangopadhyaya}
\affiliation{Theoretical High Energy Physics Division, Variable Energy Cyclotron Centre, HBNI, 1/AF Bidhannagar
Kolkata - 700064, India}

\author{Snigdha Ghosh}
\affiliation{Theoretical High Energy Physics Division, Variable Energy Cyclotron Centre, HBNI, 1/AF Bidhannagar
Kolkata - 700064, India}

\author{Sukanya Mitra}
\affiliation{Indian Institute of Technology Gandhinagar,  Gandhinagar-382355, Gujarat, India}

\author{Sourav Sarkar}
\affiliation{Theoretical High Energy Physics Division, Variable Energy Cyclotron Centre, HBNI, 1/AF Bidhannagar
Kolkata - 700064, India}

\begin{abstract}
We estimate the shear and the bulk viscous coefficients for a hot hadronic gas mixture constituting of pions 
and nucleons. The viscosities are evaluated in the relativistic kinetic theory approach by solving the transport
equation in the relaxation time approximation for binary collisions ($\pi\pi$,$\pi N$ and $NN$). 
Instead of vacuum cross-sections usually used in the literature we employ in-medium scattering amplitudes in the 
estimation of the relaxation times. The modified cross-sections
for $\pi\pi$ and $\pi N$ scattering are obtained using one-loop modified thermal propagators for $\rho$, $\sigma$ and
$\Delta$ in the scattering amplitudes which are calculated using effective interactions. The resulting suppression of 
the cross sections at finite temperature and baryon density is observed to significantly affect the $T$ and $\mu_N$ dependence 
of the viscosities of the system.

\end{abstract}

\maketitle

\section{INTRODUCTION}

During the last few decades ultrarelativistic heavy ion collision experiments
have claimed substantial attention primarily because they provide us with the opportunity to explore strongly interacting matter at
high energy densities where we can expect to find  coloured degrees of freedom in a deconfined state
known as quark-gluon plasma (QGP) in the initial stages
followed by a hot hadronic gas mixture~\cite{expt}. 
Soon after the collision, this exotic system is believed to approach a state in which the mean value of
the macroscopic quantities defining the state of the system becomes considerably larger than
their fluctuations. Transport properties have long been employed as probes to
understand the characteristics of such a thermodynamic system. The hydrodynamic evolution
of the matter created in relativistic heavy ion collision involves different dissipative
processes which can be quantified by the transport coefficients. In addition
to providing relevant insight about the dynamics of the fluid they also
carry information about how far the system is away from equilibrium.  
Recent results from RHIC have shown clear indications that the produced matter 
behaves more as a strongly interacting liquid than a weakly interacting gas.
In fact, the STAR data on elliptic flow of charged hadrons 
in Au+Au collision at $\sqrt{s}=200$ GeV per nucleon pair could be described (see e.g.~\cite{Luzum})
using very small but finite values of shear viscosity over entropy density ratio
$\eta/s$ in their viscous hydrodynamic code. A lower bound on the value
of $\eta/s(=1/4\pi)$ which follows from the uncertainty principle and substantiated using ADS/CFT correspondence~\cite{KSS}
is consistent with the values of shear viscosity extracted from
experimental results~\cite{Gavin1} and from lattice simulations~\cite{Nakamura}.
Moreover, considering the 
QCD to hadron gas transition as a cross over, $\eta/s$ shows a minimum near $T_c$, the critical temperature,
close to the lower bound mentioned~\cite{Csernai}, whereas the bulk viscosity to entropy density ratio, 
$\zeta/s$ shows large values around $T_c$~\cite{Karsch}.
 
Turning out to be an useful signature of the phase transition occurring in the medium created at RHIC
and LHC, the estimations of shear ($\eta$) and bulk ($\zeta$) viscous coefficients
have become a celebrated topic in recent times. These transport coefficients
along with their temperature behavior have been studied both below and above the 
transition temperature $T_c$. In~\cite{Thoma,Gyulassy,Baym,Heiselberg,Hosoya,AMY}
the viscosities have been estimated for interacting QCD matter employing the 
kinetic theory approach. In~\cite{JEON,Carrington,Basagoiti} shear viscosity has been obtained by evaluating
the correlation functions in the linear response theory. The bulk viscosity
has been investigated in the same spirit using Kubo formula near QCD critical point in~\cite{Moore}.
In~\cite{Sasaki,Lang,Ghosh} both the shear and bulk viscous coefficients have been estimated
employing the Nambu-Jona-Lasinio (NJL) model near chiral phase transition.
Recent quasiparticle approaches~\cite{Greco,Chandra} and the hydrodynamic simulations in~\cite{Denicol,Ryu}
have also contributed to the study of the coefficients.

In recent times a substantial amount of interest has been directed towards the analysis
of viscous coefficients in the hadronic sector of heavy ion collisions as well.
The hadron resonance gas model has been effectively applied in order to extract
the value of shear viscosity in~\cite{Noronha, Wiranata,Pal,Mishra}. 
In~\cite{Gavin,Itakura,Prakash,Davesne} the viscosities
have been estimated in the kinetic theory approach utilizing parameterized cross sections extracted from empirical data
for a hadronic gas mixture, whereas in~\cite{Purnendu,Albright} a quasiparticle model has been
used to estimate $\eta$ and $\zeta$. The NJL model also has been used to predict the 
viscous coefficient in the hadronic regime by~\cite{Buballa} using kinetic theory approach.
Scattering amplitudes evaluated using lowest order chiral perturbation theory
have been employed in~\cite{Dobado1,Cheng-Nakano} and 
a unitarized cross section has been used in~\cite{Dobado2} using the inverse amplitude method
to obtain an estimate of $\eta$. In~\cite{LuMoore} the bulk viscosity
of a pion gas has been computed including number-changing inelastic processes
using chiral perturbation theory. In~\cite{Dobado3}
the behavior of $\zeta$ has been demonstrated around the point of phase transition 
using the linear sigma model. 

The scattering cross-section that appears in the collision integral is the dynamical input in the kinetic theory approach
and it is highly suggestive that it contains the effect of the hot and/or dense medium. 
For the case of a pion gas the consequences of an in-medium cross-section on the temperature dependence of
the transport coefficients were extensively discussed in~\cite{Mitra1,Mitra2,Mitra3}. Using effective interactions and the techniques
of thermal field theory the $\pi\pi$ scattering amplitudes evaluated with self-energy corrected $\rho$ and $\sigma$ 
meson propagators in the internal lines caused a significant modification in the cross-section and consequently the viscosities~\cite{Mitra1}, thermal conductivity~\cite{Mitra2} and relaxation times of flows~\cite{Mitra3}. In view of the upcoming CBM experiment at FAIR it is thus natural to ask how the presence of a finite baryon density is likely to affect these results. To study such effects one has to include nucleons and consequently the $\pi N$ cross-section becomes a significant dynamical input in the study of transport phenomena in addition to the $\pi\pi$ cross-section.

In this work we estimate the shear and bulk viscosities of a hot and dense gas consisting of pions and nucleons. Analogous to the $\rho$ and $\sigma$ mesons mediating the $\pi\pi$ interaction we consider $\pi N$ scattering to proceed by exchange of the lightest baryon resonance, the $\De$, which is close to an 
ideally elastic $\pi N$ resonance, decaying almost entirely into pions and nucleons.  We obtain the $\De$ self-energy at finite temperature and baryon density evaluating several one-loop diagrams with $\pi$, $\rho$, $N$ and $\De$ in the internal lines using standard thermal field theoretic methods. The in-medium propagator of the $\De$ baryon is then used in the scattering amplitudes to obtain the $\pi N$ cross-section. The transport equations for the pion and nucleon  are solved to obtain the temperature and density dependence of the shear and bulk viscosities.

It is worth emphasizing at this point that whereas vacuum scattering amplitudes have been used in almost all works dealing with transport coefficients of hadronic matter discussed above, the novelty in our approach is the use of in-medium ones. This feature attributes a realistic nature to the evaluation of these quantities and makes the formalism more complete. Moreover, the use of thermal field theoretic methods ensures that scattering and decay of excitations in the medium are included systematically in the evaluation of spectral densities which play the most significant part. Incorporating the in-medium $\pi N$ cross-section in addition to the $\pi\pi$ in the coupled transport equations for the hadronic gas mixture we thus aim to provide a more reliable estimate of the viscosities, in particular their dependence on temperature and baryon density, which when used as inputs to the hydrodynamic equations will produce a more realistic scenario of space time evolution of the later stages of heavy ion collisions.

This work is organized in the following manner. Section II deals with the formulation of
the shear and bulk viscous coefficients obtained by solving the transport equation, where their
collision terms are treated in relaxation time approximation. Section III discusses 
the in-medium modification of the $\Delta$ propagator and its 
effect on the pion nucleon cross section after briefly recalling the in-medium $\pi\pi$ case. The numerical results
and corresponding discussions are presented in Section IV followed by a summary in section V. Some mathematical details are provided in appendices A, B and C.

\section{Formalism}
In order to describe the hot and dense matter created in heavy ion collisions we have two Lorentz-covariant
dynamical frameworks at our disposal; covariant transport theory and relativistic hydrodynamics.
While the evolution of macroscopic parameters related to transport processes involving thermodynamic forces and fluxes
can be studied using hydrodynamics, the microscopic collisions between constituents giving
rise to dissipative phenomena is described by the transport theory.
In linear hydrodynamic theory that satisfies the second law of thermodynamics, the relation 
between the thermodynamic forces and the corresponding fluxes is given by~\cite{Weinberg2,Degroot},
\ba
 T^{\mu\nu}=enu^{\mu}u^{\nu}-P\Delta^{\mu\nu}+\Pi^{\mu\nu}~~\nonumber\\
\Pi^{\mu\nu}=2\eta\langle\partial^{\mu}u^{\nu}\rangle +\zeta(\partial\cdotp u)\Delta^{\mu\nu}
\label{mc0}
\ea 
where $P$, $e$ and $n$ are the pressure, energy density per particle and particle density respectively. $\eta$ and $\zeta$ are the coefficients of shear and bulk viscosities, 
$u^{\mu}$ is the hydrodynamic four velocity, $\Delta^{\mu\nu}=g^{\mu\nu}-u^{\mu}u^{\nu}$ is the projection operator and $\la\cdots\ra$ denotes symmetric traceless combination. 
Throughout this paper the convention of metric that has been used is $g^{\mu\nu}=(1,-1,-1,-1)$.

  At the microscopic level energy and momentum are carried by the constituent particles and their exchange occurs due to the flow and 
collision of the particles. Viscous forces appear when the system is away from local equilibrium and work to bring the system back to equilibrium. The correspondence between kinetic theory and viscous hydrodynamics can thus be established by considering small deviation from equilibrium for which the distribution function is expressed as,
\be
f_{k}(x,p)=f^{(0)}_{k}(x,p)+\de f_{k}(x,p),~~~~\de f_{k}(x,p)=f_{k}^{(0)}(x,p)[1\pm f^{(0)}_{k}(x,p)]\phi_{k}(x,p)
\label{ff}
\ee
where the equilibrium distribution function as a function of position $x$ and momentum $p$ is given by
\be
f_k^{(0)}(x,p)=\left[\exp{\frac{p\cdot u(x)-\mu_k(x)}{T(x)}}-\sigma\right]^{-1}
\label{f0}
\ee
with $T(x)$, $u_\mu(x)$ and $\mu(x)$ representing the local temperature, 
flow velocity and chemical potential respectively. For bosonic (fermionic) degree of freedom  $\sg=1(-1)$.
The quantity $\phi_{k}$ parametrizes the deviation of the distribution function $\delta f_{k}$ of the $k^{th}$ particle specie from equilibrium. The $\pm$ sign in the expression of $\delta f_k$ denotes Bose enhancement or Pauli blocking.
The viscous part of the energy momentum tensor is then given by
\be
\Pi^{\mu\nu}=\sum_{k=1}^{N}\int d\Gamma_k\Delta^{\mu}_{\sigma}\Delta^{\nu}_{\tau} p_{k}^\sigma p_{k}^\tau\ \delta f_{k}.
\label{mc1}
\ee 
where $N$ is the number of particle species and $d\Gamma_k=d^3p_{k}/(2\pi)^3 E_{k}$ with $E_{k}=\sqrt{\vec p_{k}^2+m_{k}^2}$. On the right hand side of eq.~(\ref{mc0}) the thermodynamic forces appear with different tensorial ranks involving the space derivative of the hydrodynamic four velocity. The right hand side of eq.~(\ref{mc1}) involves integration over the particles' three momenta and in order that it conforms to the form of $\Pi^{\mu \nu}$ as expressed in
eq.~(\ref{mc0}), $\phi_{k}$ must be a linear combination of the thermodynamic forces with proper coefficients and appropriate tensorial ranks. Consequently $\phi_{k}$ is expressed as.
\be 
\phi_{k}=A_{k}\partial\cdot u-C_{k}^{\mu\nu}\langle\partial_{\mu}u_{\nu}\rangle ~,
\label{phi}
\ee
where $A_{k}$ and $C_{k}^{\mu\nu}$ are the unknown coefficients needed to be determined.
It is convenient to decompose $\Pi^{\mu\nu}$ into a traceless part and a remainder as
\be
\Pi^{\mu\nu}=\mathring{\Pi}^{\mu\nu} +\Pi \Delta^{\mu\nu}
\ee
where the viscous pressure $\Pi$ is defined as one third of the trace of the 
viscous pressure tensor,                                                                                                                                                                          
\be
\Pi=\sum_{k=0}^{N}\frac{1}{3}\int d\Gamma_k\Delta_{\sigma\tau}p_{k}^\sigma p_{k}^\tau\ \delta f_{k}.
\label{bulk1}
\ee
So the traceless part of viscous trace tensor comes out to be,
\ba
\mathring{\Pi}^{\mu\nu}&=&\Pi^{\mu\nu}-\Pi\Delta^{\mu\nu}\nonumber\\
&=&\sum_{k=0}^{N}\int d\Gamma_k\{\Delta^{\mu}_{\sigma}\Delta^{\nu}_{\tau}-
\frac{1}{3}\Delta_{\sigma\tau}\Delta^{\mu\nu}\}  p_{k}^\sigma p_{k}^\tau\ \delta f_{k}. 
\label{shear1}
\ea
 Substituting the expression for $\phi_{k}$ from (\ref{ff}) and (\ref{phi}) in eqs.~(\ref{bulk1}) and (\ref{shear1}) and comparing the coefficients with the same tensorial ranks in (\ref{mc0}) the expressions for shear and bulk viscosity turn out respectively as
 \ba  
 \eta=-\sum_{k=1}^{N}\frac{1}{10}\int d\Gamma_{k}  \langle p_{k\mu}p_{k\nu}\rangle f_{k}^{(0)}(1\pm f_{k}^{(0)})\, C_k^{\mu\nu}
 \label{eta}
 \ea  
and
 \ba 
\zeta = \sum_{k=1}^{N}\frac{1}{3}\int d\Gamma_{k}\Delta_{\mu\nu}p_{k}^\mu p_{k}^\nu &f_{k}^{(0)}(1\pm f_{k}^{(0)})&A_{k}~.
\label{zeta} 
 \ea
 
 In order to obtain the shear and bulk viscous coefficients we need to find $A_{k}$ and $C_{k}^{\mu \nu}$ for which we turn to the relativistic transport equation for a multi-particle system. This is given by
 \be 
p_{k}^\mu\partial_\mu f_{k}(x,p)=\sum_{l=1}^{N}\frac{g_{l}}{1+\delta_{kl}}C_{kl}[f_k],
\label{treq} 
 \ee
 where $g_{l}$ in the degeneracy of $l^{th}$ particle and the collision integral on the right hand side for binary elastic collisions $p_{k}+p_{l}\rightarrow p^{'}_{k}+p_{l}^{'}$ is
 \ba 
 C_{kl}[f_k]&=&\int d\Gamma_{p_{l}}\ d\Gamma_{p'_{k}}\ d\Gamma_{p'_{l}}[f_{k}(x,p'_{k})f_{l}(x,p_{l}') \{1\pm f_{k}(x,p_{k})\}
 \{1\pm f_{l}(x,p_{l})\}\nonumber\\
 &&-f_{k}(x,p_{k})f_{l}(x,p_{l})\{1\pm f_{k}(x,p'_{k})\}\{1\pm f_{l}(x,p'_{l})\}]\ W_{kl} ~.
 \label{coll}
 \ea
The dynamical input which goes into the determination of the distribution function appears in the interaction rate 
$W_{kl}=\frac{s}{2}\ \frac{d\sigma_{kl}}{d\Omega}(2\pi)^6\delta^4(p_{k}+p_{l}-p'_{k}-p_{l}')$.

 The distribution function of the particles may be expanded in a series, 
 in powers of the non-uniformity parameter (or Knudsen number, which acts as a book-keeping factor) 
 as, $f_{k}=f_{k}^{(0)}+\epsilon f_{k}^{(1)}+\epsilon ^{2}f_{k}^{(2)}+...$ and 
 substituted in the multi-particle transport equation. Restricting to the first order the transport equation becomes~\cite{Degroot},
 \be 
 p^{\mu}u_{\mu}Df^{(0)}_{k}+p^{\mu}\nabla _{\mu}f^{(0)}_{k}=\sum_{l=1}^{N}\frac{g_{l}}{1+\delta_{kl}}C_{kl}[f_k^{(1)}]
 \label{teq2}
 \ee   
 where the derivative on the left hand side is separated into a time-like and a space-like part using $\del_\mu=u_\mu D+\nabla _{\mu}$ 
 where $D=u^{\nu}\partial_{\nu}$ and $\nabla _{\mu}=\nabla_{\mu \nu}\partial^{\nu}$. Using the conservation equations the time derivative is replaced with space derivatives of the fluid four-velocity so that the left hand side of eq.~(\ref{teq2}) becomes
 \be 
 \frac{1}{T}{f_{k}^{(0)}(1\pm f_{k}^{(0)})}[Q_{k}\partial\cdot u-
 \langle p_{k}^{\mu}p_{k}^{\nu} \rangle \langle \del_{\mu} u_{\nu}\rangle]
 \label{relax_left}
 \ee
 where  $Q_{k}=T^2[-\frac{1}{3}z_{k}^2+\tau_{k}^2(\frac{4}{3}-\gamma)+\tau_{k}
 \{(\gamma_{k}''-1)\hat{h}_{k}-~\gamma_{k}'''\}]$,
 $z_{k}=\frac{m_k}{T}$, $\tau_{k}=\frac{(p_{k}\cdot u)}{T}$ and $\hat{h}_{k}=\frac{h_k}{T}$
 with $h_k$ is the enthalpy per particle belonging to $k^{th}$ species. The 
 details of the calculation along with the expressions of
 $\gamma$'s and $z$'s are given in Appendix-B.
 The bulk and shear viscous forces are represented
 by the first and second terms respectively in (\ref{relax_left}) where we have ignored terms related to thermal conductivity and diffusion.
 
 The term on the right hand side of (\ref{treq}) is now simplified by assuming that all particles involved in the 
 scattering process except the $k^{th}$ particle with momentum $p_{k}$ is in equilibrium. This is the so-called 
 relaxation time approximation. In the collision integral given by eq.~(\ref{coll}) we thus use 
 $f_{k}^{(0)}+\delta f_{k},~f_{l}^{(0)},~f_{k}^{(0)'}$ and $f_{l}^{(0)'}$ in place of $f^{(1)}_{k},~f_{l}^{(1)},~f_{k}^{(1)'}$ and $f_{l}^{(1)'}$ 
 respectively to get
\be 
\sum_{l=1}^{N}\frac{g_{l}}{1+\delta_{kl}}C_{kl}[f_k]=-\frac{\delta f_k}{\tau_k}E_{k}
\ee
where the relaxation time is given by
\be 
[\tau_k(p_k)]^{-1}=~\sum_{l=1}^{N}[\tau_{kl}(p_k)]^{-1}
\ee
with
\be
[\tau_{kl}(p_k)]^{-1}=\frac{g_l}{1+\delta _{kl}}\frac{{\rm csh}(\epsilon_{k}/2)}{E_k}\int d\omega _{l}d\omega _{k}^{'}d\omega _{l}^{'}W_{kl}
\ee
where $d\omega _k=d\Gamma_{p_k}/[2~{\rm csh}(\epsilon_{k}/2)]$ , $\epsilon_k=(E_{k}-\mu_{k})/T$ and the function\\
${\rm csh}(x_k)=\cosh(x_k)$ if $k$ represents a fermion and ${\rm csh}(x_k)=\sinh(x_k)$ if $k$ is a boson.

The transport equation (\ref{treq}) in the relaxation time approximation thus takes the form
 \be 
 \frac{1}{TE_k}[Q_{k}\partial\cdot u-
 \langle p_{k}^{\mu}p_{k}^{\nu} \rangle \langle \del_{\mu} u_{\nu}
 \rangle]=-\frac{\phi_k}{\tau_k}
 \label{relax1}
 \ee
where use has been made of (\ref{ff}). 
 Substituting the expression of $\phi_k$ from eq.~(\ref{phi}) in eq.~(\ref{relax1}) and comparing the coefficients of 
 independent thermodynamic forces on both sides, we obtain the coefficients $A$ and $C^{\mu\nu}$ in terms of $\tau_k$,
 \ba
 A_{k}=&&-\frac{\tau_k}{TE_k}Q_k~,\\
 C_{k}^{\mn}=&&-\frac{\tau_k}{TE_k}\langle p_{k}^{\mu}p_{k}^{\nu} \rangle~.
 \ea
From eq.~(\ref{eta}) and (\ref{zeta}), using the conditions mentioned in Appendix-C, we finally arrive at the expressions of shear and bulk viscosity,
 \ba
 \eta=&&\frac{1}{15T}\sum_{k=1}^{N}\int \frac{d^3 p_{k}}{(2\pi)^3}\frac{\tau_k}{E_k^2}|\vec{p_k}|^4 {f_{k}^{(0)}(1\pm f_{k}^{(0)})}~,\\
 \zeta=&&\frac{1}{T}\sum_{k=1}^{N}\int \frac{d^3 p_{k}}{(2\pi)^3}\frac{\tau_k}{E_k^2}\{Q_k^2\} {f_{k}^{(0)}(1\pm f_{k}^{(0)})}~.
 \ea

\section{Dynamical Inputs}

\begin{figure}
 \includegraphics[scale=0.4,angle=-90]{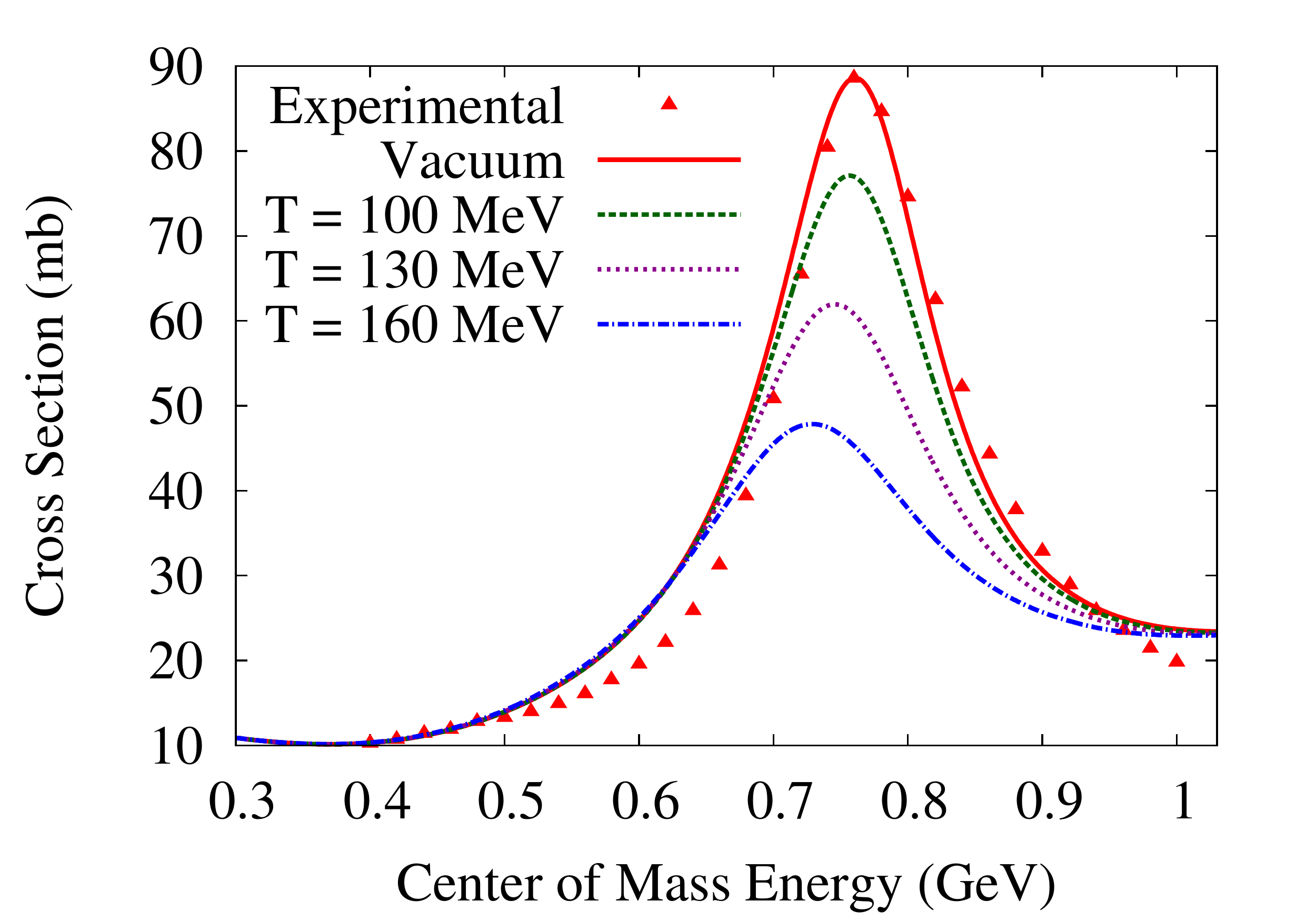}
 \caption{The $\pi\pi$ cross-section as a function of $E_{c.m.}$. }
 \label{pipisig}
 \end{figure}
We now specialize to the case of a pion nucleon gas mixture so that the index $k=\pi,N$. The relaxation times of pions and nucleons are thus respectively given by
\ba
\tau_\pi^{-1}&=&\tau_{\pi\pi}^{-1}+\tau_{\pi N}^{-1}\nn\\
\tau_N^{-1}&=&\tau_{\pi N}^{-1}+\tau_{N N}^{-1}~.
\label{rel_piN}
\ea

Here we aim to demonstrate the effects of the in-medium cross-sections on the viscosities. Correspondingly, we will first obtain the $\pi\pi$ and $\pi N$ cross-sections in the medium. The case of $\pi\pi$ has been discussed extensively in~\cite{Mitra1}. We however recall the main results for completeness. We assume the $\pi\pi$ scattering to proceed via $\rho$ and $\sigma$ meson exchange. Using the interaction $\cl=g_\rho\vec\rho^\mu\cdot\vec \pi\times\del_\mu\vec \pi+\frac{1}{2}g_\sigma
m_\sigma\vec \pi\cdot\vec\pi\sigma$
with $g_\rho=6.05$ and $g_\sigma=2.5$, the matrix elements in the isoscalar and isovector channels are given by 
\ba
\cm_{I=0}&=&2g_\rho^2\left[\frac{s-u}{t-m_\rho^2}+\frac{s-t}{u-m_\rho^2}\right]\nonumber\\
&+&g_\sigma^2 m_\sigma^2\left[\frac{3}{s-m_\sg^2+\Sg_\sg}+\frac{1}{t-m_\sg^2}+
\frac{1}{u-m_\sg^2}\right]\nonumber\\
\cm_{I=1}&=&g_\rho^2\left[\frac{2(t-u)}{s-m_\rho^2+\Sg_\rho}+
\frac{t-s}{u-m_\rho^2}-\frac{u-s}{t-m_\rho^2}\right]\nonumber\\   
&+&g_\sigma^2 m_\sigma^2\left[\frac{1}{t-m_\sg^2}-\frac{1}{u-m_\sg^2}\right]~.
\label{amp}
\ea
where the corresponding $s$-channel propagators have been replaced by effective ones obtained by a Dyson-Schwinger sum of one-loop self-energy diagrams in vacuum. The cross-section is given by $\sigma=\frac{1}{64\pi^2 s}\int\overline{|\cm|^2}d\Omega$ 
where the isospin averaged amplitude is defined as $\overline{|\cm|^2}=\frac{1}{9}\sum(2I+1)\overline{|\cm_I|^2}$. As seen in fig.~\ref{pipisig} the cross-section agrees fairly well with the experimental values up to about 1 GeV. In the medium the vacuum self energies $\Sg_\rho$ and $\Sg_\sigma$ are replaced with in-medium ones evaluated using thermal field theory~\cite{Sourav_RT,Bellac}. For the $\sigma$ meson only
the $\pi\pi$ loop graph is evaluated in the medium
whereas in case of the $\rho$ meson in addition to the $\pi\pi$ loop diagram,
$\pi\omega$, $\pi h_1$, $\pi a_1$ self-energy diagrams
are included~\cite{Sabya}. The imaginary part of the self-energy is given by
\ba
&&{\rm Im} \Sg(q_0,\vq)=-\pi\int\frac{d^3 k}{(2\pi)^3 4\om_\pi\om_h}\times\nonumber\\
&& \left[L_1(1+n_+(\om_\pi)+n_+(\om_h))\de(q_0-\om_\pi-\om_h)\right.\nonumber\\
&& +\left. L_2(n_-(\om_\pi)-n_+(\om_h))\de(q_0+\om_\pi-\om_h)\right]
\label{ImPi_a}
\ea
where $n_\pm(\om)=\frac{1}{e^{(\om\mp\mu)/T}-1}$ is the Bose distribution
function with arguments $\om_\pi=\sqrt{\vk^2+m_\pi^2}$ and
$\om_h=\sqrt{(\vq-\vk)^2+m_h^2}$. The terms $L_1$ and $L_2$ arise from
factors coming from the vertex etc, details of which 
can be found in~\cite{Sabya}. The angular integration is
done using the $\de$-functions which 
define the kinematic domains for occurrence of scattering and decay processes
leading to loss or gain of $\rho$ (or $\sigma$) mesons in the medium. The term with $L_1$ arises from the unitary cut and corresponds to formation and decay in the medium weighted by Bose enhancement factors and the second term corresponds to the so-called Landau cut contribution arising from resonant scattering in the medium.
To account for the substantial $3\pi$ and $\rho\pi$ branching ratios of some of the unstable
particles in the loop the self-energy function is convoluted with their spectral functions~\cite{Sabya}. 
The increase of the imaginary part for $T=$130 and 160 MeV is manifested in a suppression of the magnitude of the cross-section as seen in fig.~\ref{pipisig}. 

\begin{figure}
 \includegraphics[scale=0.4,angle=-90]{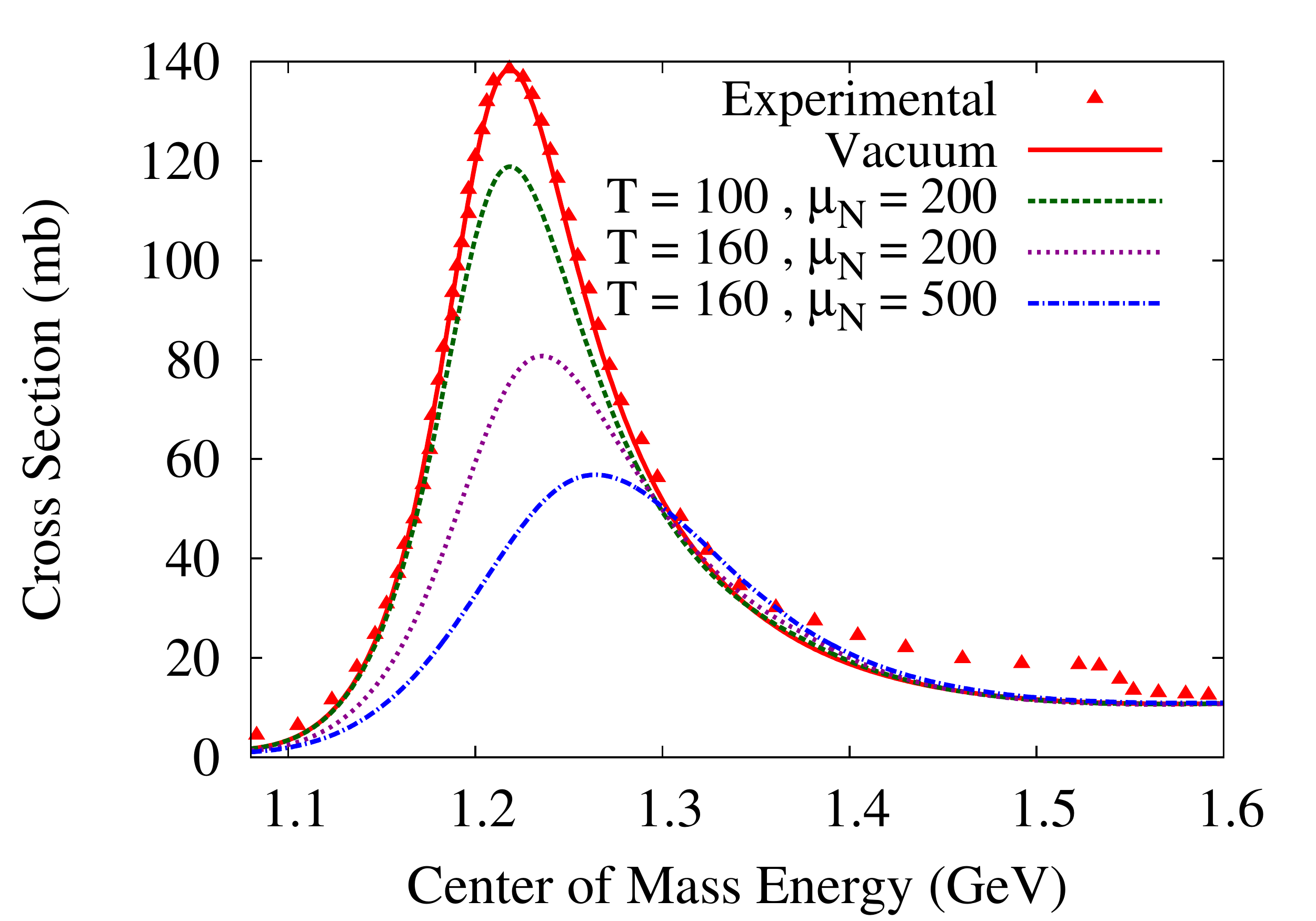}
 \caption{The $\pi N$ cross-section as a function of $E_{c.m.}$. The values of temperature $T$ and nucleon chemical potential $\mu_N$ are in MeV.}
 \label{delsig}
 \end{figure}
The case of $\pi N$ scattering is treated analogously. It is taken to proceed via the exchange of the $\De$-baryon which is the lightest baryon resonance. Despite being relatively broad, it is well separated from other resonances. We use the well-known interaction $\mathcal{L}_{\pi N\Delta} = \frac{f_{\pi N\Delta}}{m_\pi}\bar{\De}^\mu\vec T^\dagger\partial_\mu\vec{\pi}\psi + H.c.$ with $f_{\pi N\Delta}=2.8$ to evaluate the scattering matrix elements.
Averaging over isospin, the squared invariant amplitude for the process $\pi(k)~N(p)\rightarrow~\pi(k')~N(p')$
is given by
\ba
\overline{|\mathcal{M}|^2}
&=&\frac{1}{3}\left(\frac{f_{\pi N\Delta}}{m_\pi}\right)^4\left[ \frac{F^4(k,p)T_s}{\left|s-m_\Delta^2-{\Pi}\right|^2} + \frac{F^4(k,p')T_u}{\left(u-m_\Delta^2\right)^2}\right. \nn\\
&&\hskip 2.5cm\left.+ \frac{2F^2(k,p)F^2(k,p')T_m (s-m_\Delta^2-\text{Re}{\Pi})}{3(u-m_\Delta^2)\left|s-m_\Delta^2-{\Pi} \right|^2} \right]
\label{modmsq}
\ea
where $T_s$, $T_u$ and $T_m$ stand for
\begin{eqnarray}
T_s &=& \text{Tr}\left[(\cancel{p}'+m_N)D_s(\cancel{p}+m_N)\gamma^0D_u^\dagger\gamma^0 \right] \\
T_u &=& \text{Tr}\left[(\cancel{p}'+m_N)D_u(\cancel{p}+m_N)\gamma^0D_u^\dagger\gamma^0 \right] \\
T_m &=& \text{Tr}\left[(\cancel{p}'+m_N)D_s(\cancel{p}+m_N)\gamma^0D_u^\dagger\gamma^0 \right] 
\end{eqnarray}
in which
\begin{eqnarray}
D_s &=& k_\alpha k'_\beta\mathcal{O}^{\beta\nu}\Sg_\mn(q_s) \mathcal{O}^{\mu\alpha} \\
D_u &=& k'_\alpha k_\beta\mathcal{O}^{\beta\nu}\Sg_\mn(q_u) \mathcal{O}^{\mu\alpha}
\end{eqnarray}
and 
\[
\Sigma_{\alpha\beta}(q)=\left(\cancel{q}+m_q\right)\left[ -g_{\alpha\beta}+\frac{1}{3m_q^2}q_\alpha q_\beta + \frac{1}{3}\gamma_\alpha\gamma_\beta+\frac{1}{3m_q}\left( \gamma_\alpha q_\beta-\gamma_\beta q_\alpha \right) \right]~.
\]
At each vertex we consider the form factor~\cite{Snigdha}
\be 
F(p,k)=\frac{\Lm^2}{\Lm^2+(\frac{p\cdot k}{m_p})^2-k^2}
\ee
in which $p$ and $k$ denote the momenta of the fermion and boson respectively. The cut-off is taken as $\Lm=600$ MeV~\cite{Snigdha}. As seen from fig.~\ref{delsig} the one-loop $\De$ self-energy in vacuum produces a good description to the experimental $\pi N$ scattering cross-section. At finite temperature we consider additional contributions coming from on-shell particles in the medium by evaluating $\pi N$,  $\rho N$, $\pi \De$ and $\rho\De$ self-energies using the real time method. The expression for the spin averaged imaginary part of self-energy is given by
\begin{eqnarray}
\text{Im}{\Pi} =&& -\pi\int\frac{d^3k}{(2\pi)^3}\frac{1}{4\omega_k\omega_p}\nn\\
&&\left[N_1(1+n_+(\omega_k)-\tilde{n}_+(\omega_p))\delta(q_0-\omega_k-\omega_p)\right.\nn\\ 
&&+\left. N_2(n_-(\omega_k)+\tilde{n}_+(\omega_p))\delta(q_0+\omega_k-\omega_p)\right]
\label{im1}
\end{eqnarray} 
where the distribution function for the fermions is given by $\tilde{n}_\pm(\omega)=\frac{1}{e^{\beta\left(\omega\mp\mu\right)}+1}$.
The expressions for $N_1$ and $N_2$ for the different loops may be found in~\cite{Snigdha}.  As before the first term is the contribution from decay and formation of the $\De$ baryon weighted by thermal factors. The second term is a result of scattering processes in the medium leading to the absorption of the $\De$. These processes contribute significantly to the imaginary part which is reflected as a suppression of the $\pi N$ cross-section. The suppression increases with increasing temperature and chemical potential as seen in fig.~\ref{delsig}.

\section{Results}

\begin{figure}
 \includegraphics[scale=0.4]{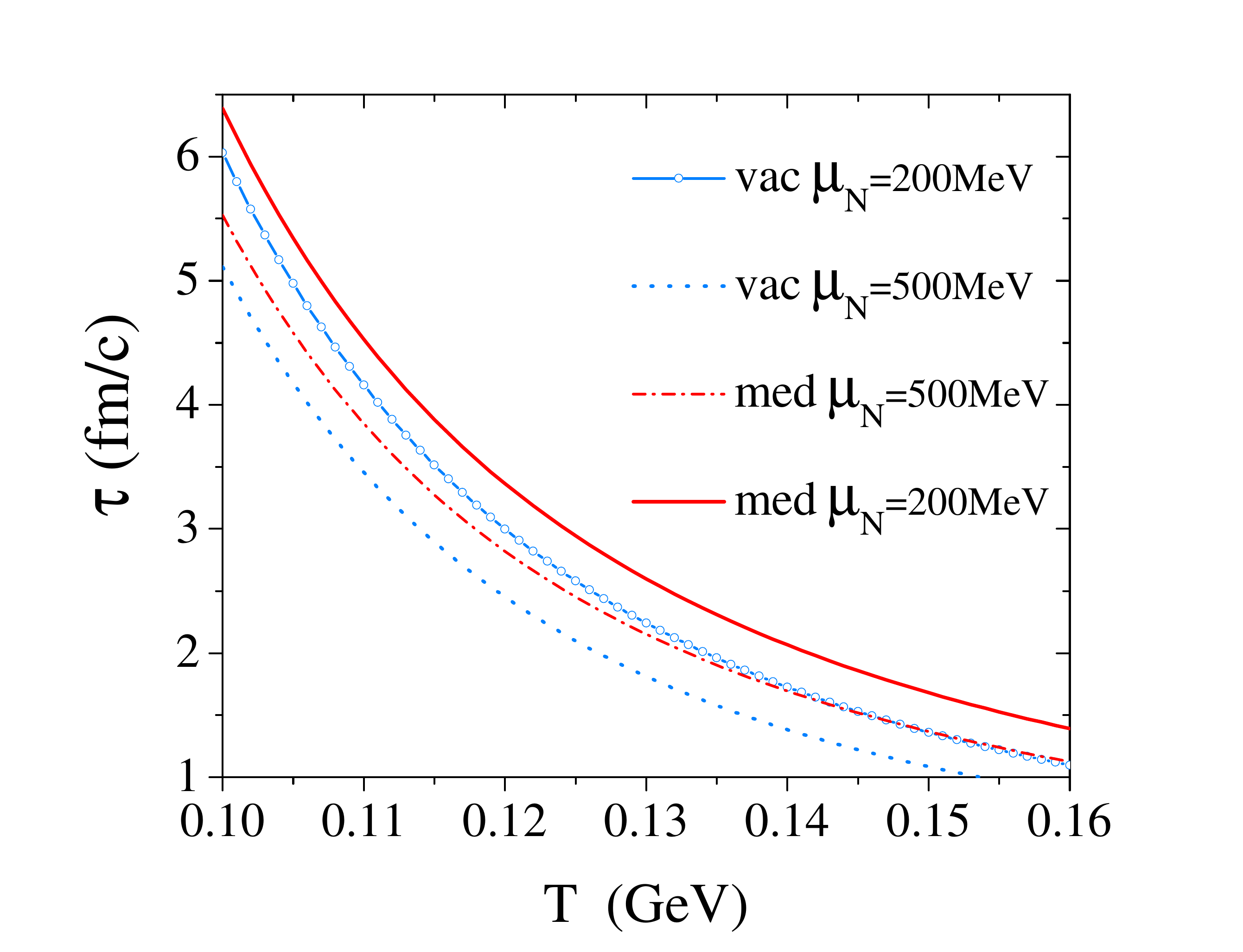}
 \caption{The relaxation time of pions in a hadron gas mixture of pions and nucleons. }
 \label{Pi_relax}
 \end{figure}

In this section we show how the medium modified $\pi\pi$ and $\pi N$ cross-sections discussed above are reflected in the relaxation times and consequently 
in the viscosities of the system. We will show results for the temperature range 100 to 160 MeV which is typical of a hadron gas produced 
in the later stages of heavy ion collisions between kinetic and chemical freeze-out. Accordingly, we consider a non-zero value of the 
pion chemical potential~\cite{Bebbie} in addition to the chemical potential for nucleons.

We plot in  fig.~\ref{Pi_relax} the relaxation time of pions in the hadron gas consisting of pions and nucleons as a function 
of temperature for $\mu_N=$200 and 500 MeV and $\mu_\pi=$80 MeV~\cite{Hirano}. The order of magnitude
of $\tau$ along with its decreasing trend with increasing temperature is consistent with \cite{Prakash}. As can be seen from 
(\ref{rel_piN}) it's magnitude is 
decided by both $\pi \pi$ and $\pi N$ cross-sections. As discussed above, both of these decrease in the medium causing the in-medium 
relaxation time of pions to effectively increase.   
 
\begin{figure}
 \includegraphics[scale=0.4]{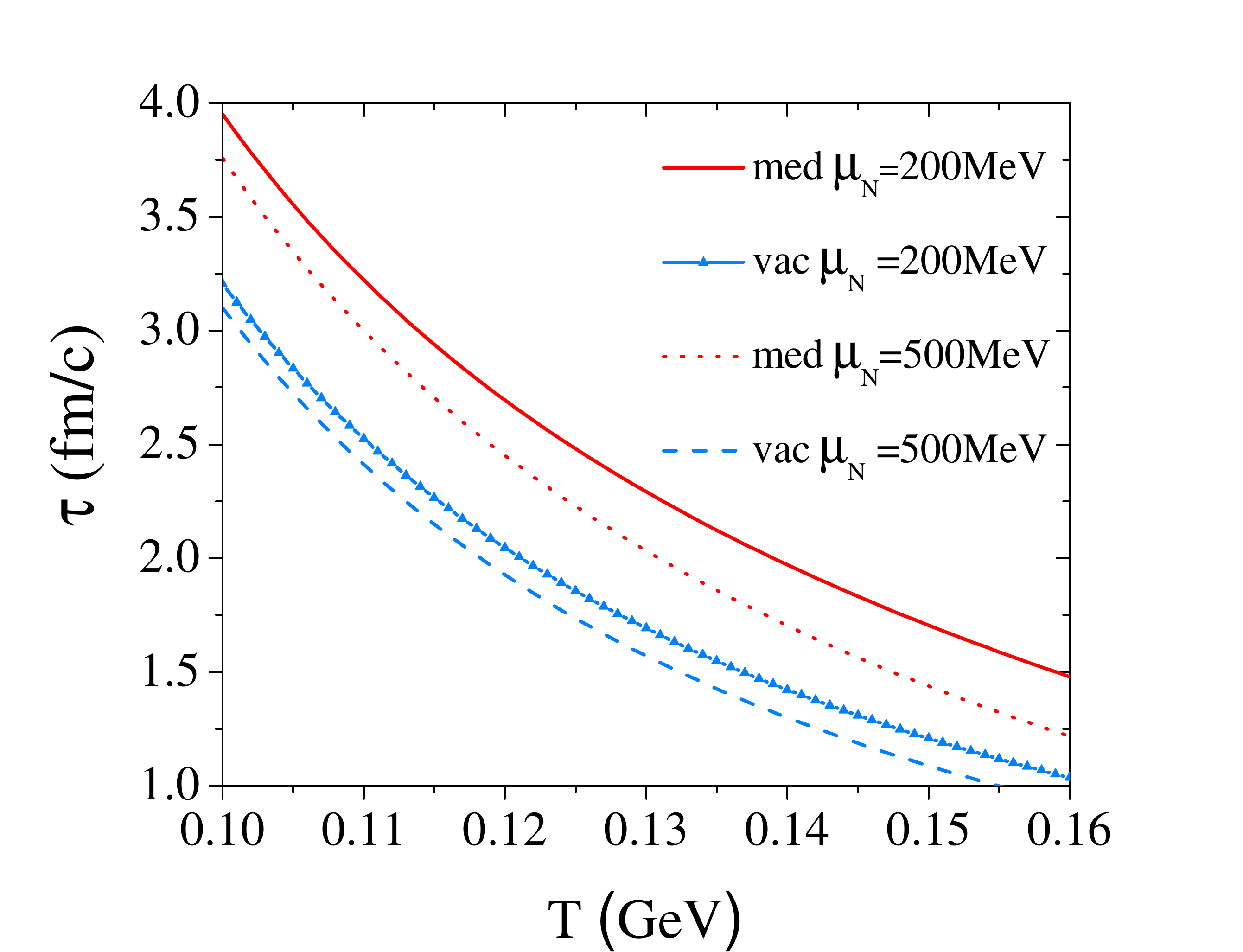}
 \caption{The relaxation time of nucleons in a hadron gas mixture of pions and nucleons. }
 \label{N_relax}
 \end{figure}

Similar features are observed in fig.~\ref{N_relax} where the relaxation time of nucleons is plotted as a function of $T$. 
The medium effect arising from the reduced $\pi N$ cross-section because of the additional scattering
and decay processes in this case cause an increase in the nucleon relaxation time.

\begin{figure}
\includegraphics[scale=0.4]{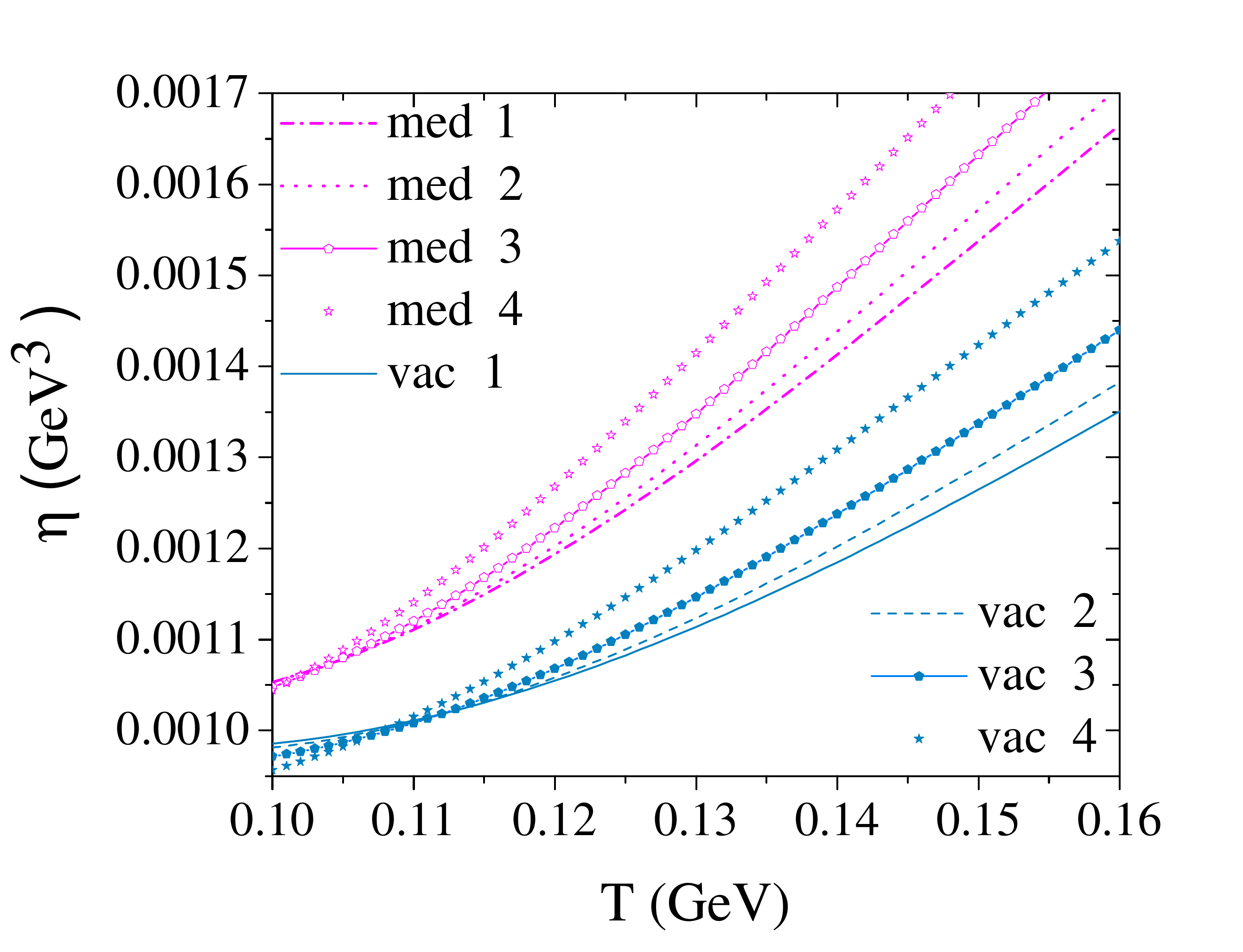}
\caption{Shear viscosity as a function of $T$ for various $\mu_N$ with and without medium effects. Legends 1,2,3 and 4 indicate $\mu_N=$200, 300, 400 and 500 MeV respectively. The pion chemical potential $\mu_\pi=80$ MeV. }
\label{Shear}
\end{figure}
\begin{figure}
\includegraphics[scale=0.4]{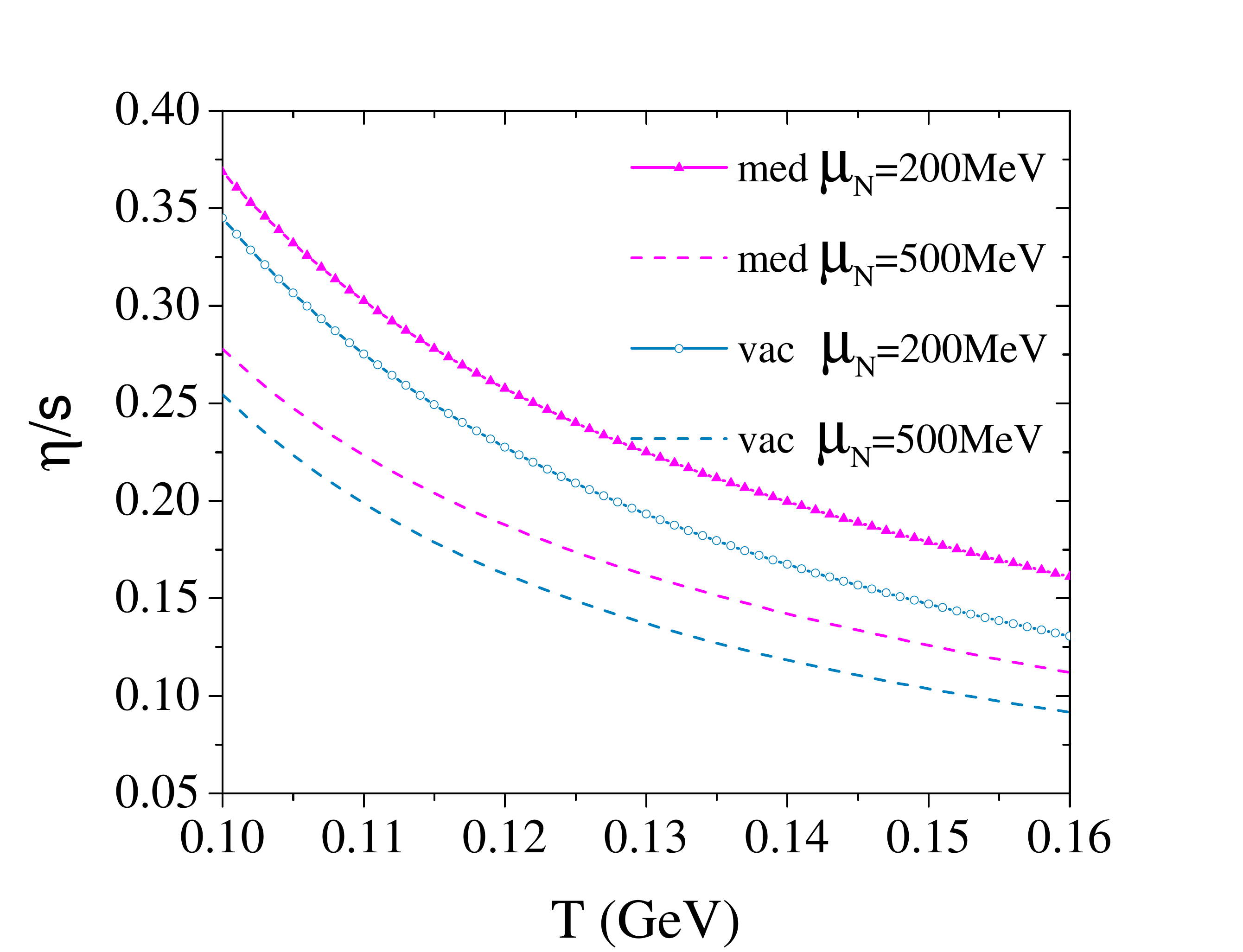}
\caption{Shear viscosity over entropy density as a function of $T$. }
\label{Shear_s}
\end{figure}
We now look into the behavior of the viscous coefficients for different values of parameters evaluated numerically. The shear viscous coefficient as
 a function of temperature is depicted in fig.~\ref{Shear}. The lower set of curves are the ones evaluated using the vacuum cross sections, while the upper set is  
 evaluated taking into consideration the medium effect on the $\pi\pi$ and $\pi N$ cross sections. The different curves in each set correspond to different 
 values of the nucleon chemical potential $\mu _{N}$ while the pion chemical potential $\mu_{\pi}$ is taken to be 80 MeV. Both in vacuum and medium the shear viscosity appears to increase with increasing $\mu_N$ and is the result of interplay of various factors. This was already noted in~\cite{Itakura} where by means of a simplified estimate of the viscosity of the mixture this feature could be understood as resulting from an enhancement of the nucleon component with increasing $\mu_N$. Moreover, a considerable change in the value of the shear viscous coefficient is seen due to the introduction of the medium effect. The decrease of the in-medium cross-section with increasing $T$ and $\mu_N$ results in an increase in the relaxation time and hence the viscosity as shown by the upper set of curves.

\begin{figure}
\includegraphics[scale=0.4]{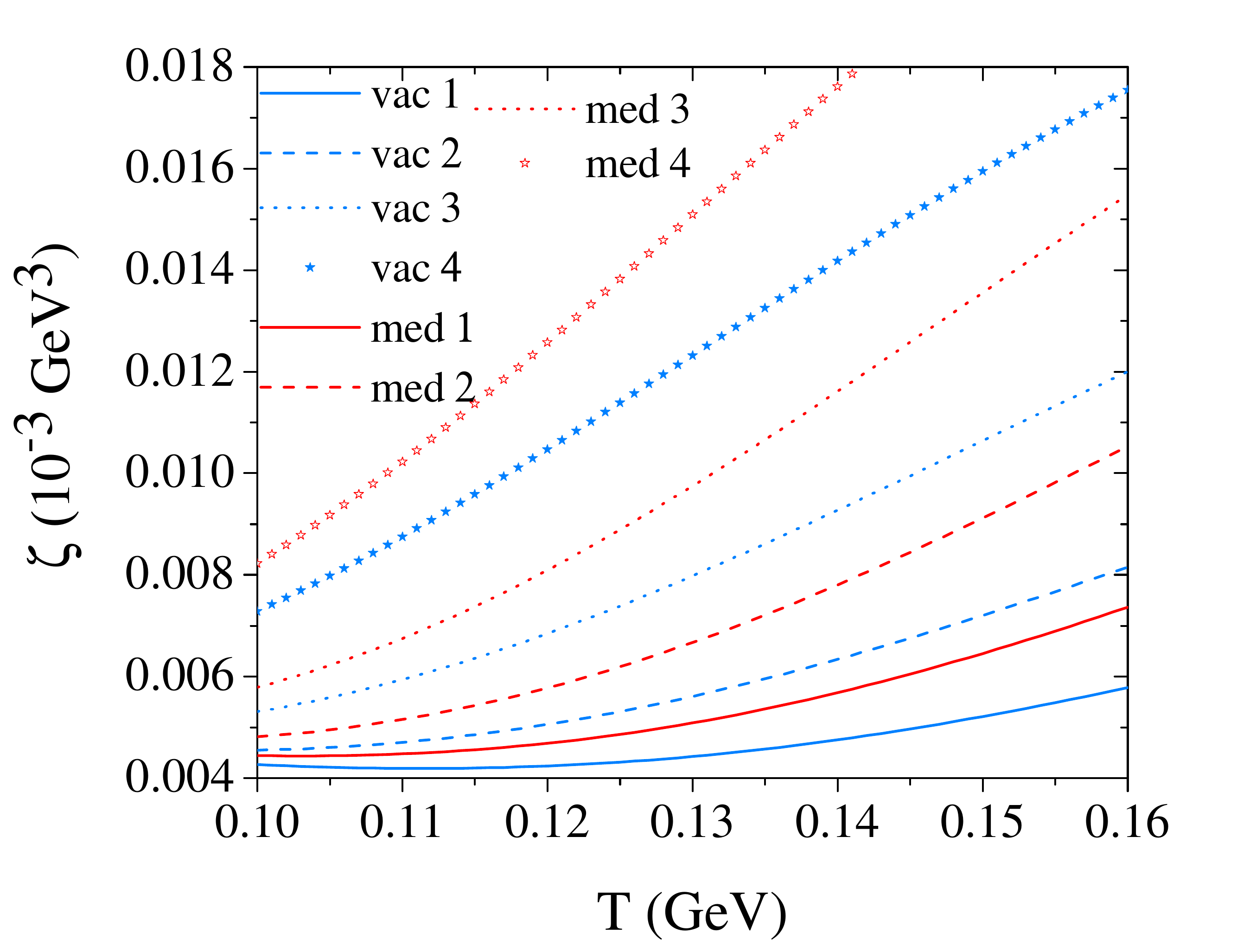}
\caption{The bulk viscosity as a function of $T$ with and without medium effects. Legends 1,2,3 and 4 indicate $\mu_N=$200, 300, 400 and 500 MeV respectively.}
\label{Bulk}
\end{figure}

 We also find a significant influence of the medium on the kinematic shear viscosity (i.e,$\eta/s$; $s$ is the entropy density of the system).
 Fig.~\ref{Shear_s} depicts the variation of kinematic viscosity with temperature for nucleon chemical potential $200$ MeV and $500$ MeV 
 represented by lines with and without symbols respectively. The lower curve in each set corresponds
to vacuum cross section while the higher one corresponds to the one calculated taking into account the medium effect for the same nucleon chemical potential. The monotonous decrease in $\eta/s$ with $T$ may be attributed to the increase in the entropy density with $T$~\cite{Itakura}. The decrease 
however, respects the lower bound~\cite{KSS} around the transition temperature.

\begin{figure}
\includegraphics[scale=0.4]{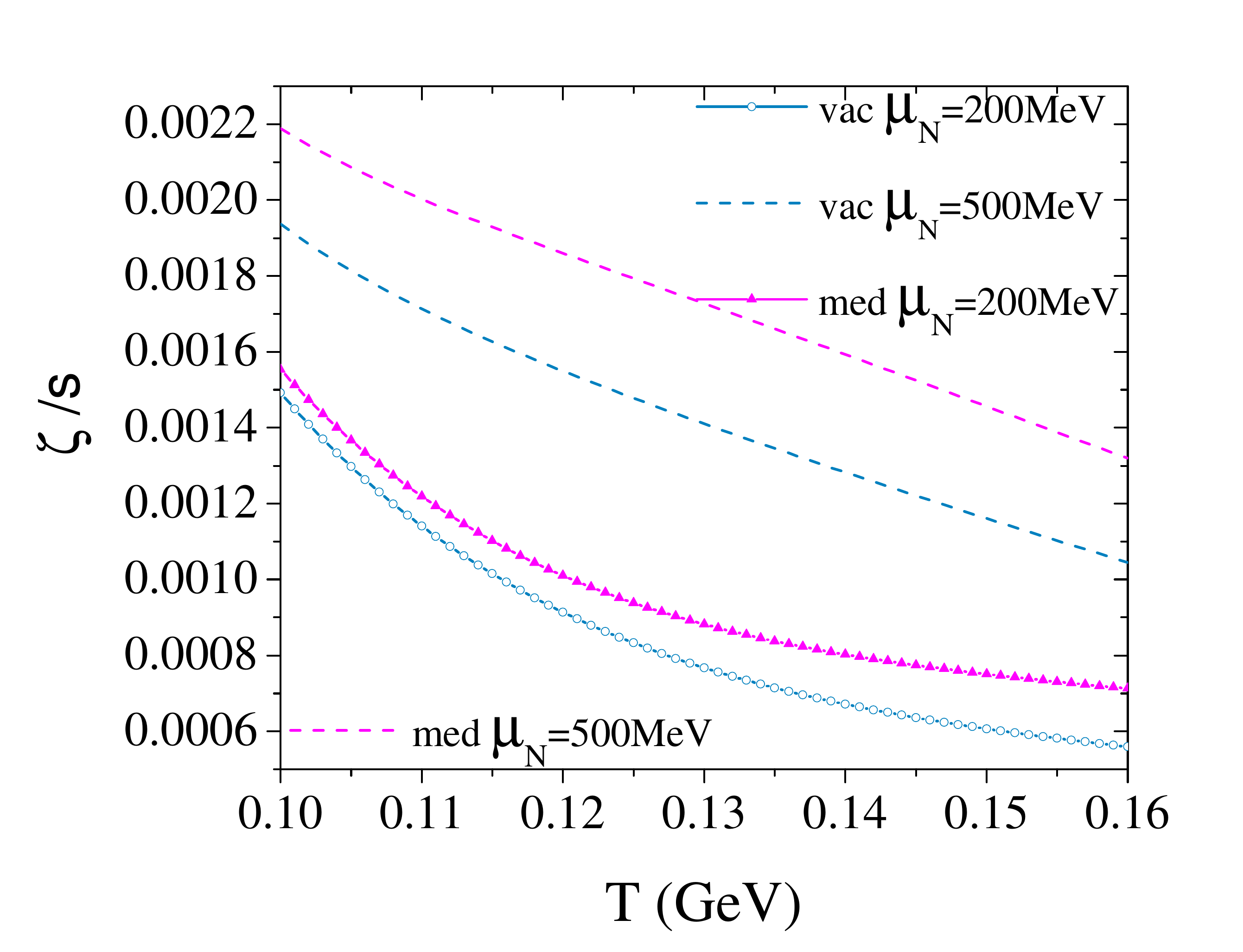}
\caption{Ratio of bulk viscosity and entropy density vs $T$.}
\label{Bulk_s}
\end{figure}

 Now we turn to the bulk viscous coefficient whose temperature dependence is shown in fig.~\ref{Bulk}. Proceeding from the bottom of the 
 diagram each pair of graphs corresponds to a different value of nucleon chemical potential. The lower curve for each pair corresponds 
 to the one where vacuum cross section has been used to calculate the relaxation time while the upper curve corresponds to the one where the medium effects have been taken into consideration. Although the 
 magnitude of bulk viscosity appears to be much smaller that that of shear viscosity the effects of the 
 thermal medium are quite visible in this case too.
 In fig.~\ref{Bulk_s}  the variation of the kinematic bulk viscosity (i.e, $\zeta/s$) has been studied where the lines correspond to the same combination of parameters as in fig.~\ref{Shear_s}. The dependence with $T$ and $\mu_N$ in vacuum and medium show similar features as the shear viscosity.  Significant difference is observed between the magnitude of viscosities calculated with vacuum and medium cross-sections.
 
\section{Summary}

In this work we evaluate the shear and bulk viscosities of a hadronic gas mixture consisting of pions and nucleons. We aim to study the difference
caused by the use of in-medium cross-sections in the transport equations. Analogous to the $\pi\pi$ scattering amplitudes which have been modified through self-energy corrections of the exchanged $\rho$ and $\sigma$ mesons, the $\pi N$ interaction at finite temperature and baryon density has been obtained by means of loop corrections to the $\De$ propagator using the techniques of thermal field theory. Because of a significant contribution to the imaginary part basically coming from (resonant) scatterings of the exchanged $\De$ baryon in the medium there is a significant suppression of the cross-section around the peak position. This causes the relaxation times to increase with respect to their vacuum values which in turn bring in a quantitative change in the viscosities  of the system. With such input in the viscous hydrodynamic simulations, the space-time evolution of the later stages of heavy ion collisions might be observably affected.

\section*{Appendix-A}
The quantities like energy density, pressure, entropy etc. of the system consisting of pions and nucleons are expressed in terms of $S_n^{\alpha}\left(z_{\pi}\right)$ and $T_n^{\alpha}\left(z_{N}\right)$, where $z_{\pi}=m_{\pi}/T$ and $z_N=m_{N}/T$. They appear as
\ba
n_{\pi}&=&g_{\pi}\int d\Gamma_{p_{\pi}} E_{p_{\pi}}
f_{\pi}^{(0)}(p_{\pi})=\frac{g_\pi}{2\pi^2}z_{\pi}^2T^3S_2^1(z_{\pi}),\nn\\
P_{\pi}&=&g_{\pi}\int
 d\Gamma_{p_{\pi}}\frac{\vp_{\pi}^2}{3}f_{\pi}^{(0)}(p_{\pi})=\frac{g_\pi}{2\pi^2}z_{\pi}^2T^4S_2^2(z_{\pi}),\nn\\
n_{\pi}e_{\pi}&=&g_{\pi}\int d\Gamma_{p_{\pi}} E_{p_{\pi}}^2f_{\pi}^{(0)}(p_{\pi})
=\frac{g_{\pi}}{2\pi^2}z_{\pi}^2T^4[z_{\pi} S_3^1(z_{\pi})-S_2^2(z_{\pi})]\nn\\
n_{\pi}h_{\pi}&=&n_{\pi}z_{\pi}T\frac{S_3^1(z_{\pi})}{S_2^1(z_{\pi})}~,
\ea
where $E_p=\sqrt{p^2+m^2}$, $f_{\pi}^{(0)}(p_{\pi})=[e^{\beta(E_{p_\pi}-\mu_{\pi})}-1]^{-1}$
and $d\Gm_p=d^3p/(2\pi)^3E_p$. Using the formula $[a-1]^{-1}=\displaystyle\sum_{n=1}^\infty(a^{-1})^n$ the function $f_{\pi}^{(0)}$ has been expanded, hence the integrals was expanded to a sum of integral which can be compactly expressed as $S_n^\alpha(z_{\pi})=\displaystyle\sum_{k=1}^\infty e^{k\mu_{\pi}/T} k^{-\alpha} K_n(kz_{\pi})$ , $K_n(x)$ denoting the modified Bessel function of order $n$ given by
\be
K_n(x)=\frac{2^n n!}{(2n)!\ x^n}\int_x^\infty\ d\tau
(\tau^2-x^2)^{n-\frac{1}{2}}e^{-\tau}
\ee
or
\be
K_n(x)=\frac{2^n n!(2n-1)}{(2n)!\ x^n}\int_x^\infty\ \tau d\tau
(\tau^2-x^2)^{n-\frac{3}{2}}e^{-\tau}~.
\ee
In the corresponding expressions for $n_N$,$P_N$,$e_N$ etc, the $S^{\alpha}_n(z_{\pi})$ will be replaced by  $T^{\alpha}_n(z_{N})$ defined as  $T_n^\alpha(z_{N})=\displaystyle\sum_{k=1}^\infty \left(-1\right)^{k-1}e^{k\mu_{N}/T} k^{-\alpha} K_n(kz_{N})$.

\section*{Appendix-B}
 The left hand side of the linearized transport equation for each species
 \be 
p^{\mu}\del_\mu f^{(0)}_{k}= p^{\mu}u_{\mu}Df^{(0)}_{k}+p^{\mu}\nabla_{\mu}f^{(0)}_{k}=-\frac{\delta f_k}{\tau_k}E_{k}.
 \label{apC1}
 \ee  
 is to be expressed in terms of thermodynamic forces. In order to do this the derivative on the left is expressed in terms of the derivatives of the thermodynamic parameters getting  
 \be
 (p_{k}\cdot u)\left[\frac{p_{k}\cdot u}{T^2}DT+D\left(\frac{\mu _{k}}{T}\right)-\frac{p^\mu_{k}}{T}
 Du_\mu\right]+p^\mu\left[\frac{p_{k}\cdot
 u}{T^2}\nabla_\mu T+\nabla_\mu \left(\frac{\mu _{k}}{T}\right)
 -\frac{p_{k}^\nu}{T}\nabla_\mu u_\nu\right]=-\frac{\delta f_k}{\tau_k}E_{k}~.
 \label{apC2}
 \ee
 Note $D \to \del /\del t$ and $\nabla_\mu \to \delta_{i}$ which are the time derivative and the space derivative respectively in the local rest frame.
 The thermodynamic forces do not contain terms like $DT$ and $D\left(\frac{\mu_{k}}{T}\right)$. Rather, it contains space derivatives of temperature, chemical potential and the thermodynamic velocity $u^{\mu}$. In order to express the time derivative in terms of space derivative of the thermodynamic parameters we make use of the conservation equations which are given up to first order by
 \ba 
  \del_\mu N^\mu_{k}=0~,\nonumber\\
  Dn_{k}=-n_{k}\del_\mu u^\mu ~.
  \label{apC3}
  \ea 
  and
\ba 
\del_\nu T^{\mn\, (0)}=0~,~~~~~~~~~~~~~~~~~~~~~~~~~    u_\mu\del_\nu T^{(0)\,  \mn}=0\nonumber\\
De=-\frac{P}{n}\del_\mu u^\mu~,~~~~~~~~\sum_{k}n_{k}De_{k}=-\left(\sum_{k}P_{k}\right)\del_\mu u^\mu
\label{apC4}
\ea  
 where $N_{k}^{\mu}=n_{k}u^{\mu}$ and $T^{\mu \nu}=n\left[(e+\frac{P}{n})u^\mu u^\nu-g^\mn \frac{P}{n}\right]+T^{(1)\, \mu \nu}$; 
 $T^{(1)\, \mu \nu}=\left(I^{\mu}_{q}u^{\nu}+I^{\nu}_{q}u^{\mu}\right)+~\Pi^{\mu \nu}$. The quantities $I_{q}^{\mu}$ and $\Pi^{\mu \nu}$ are the heat flow and the viscous part of the energy momentum tensor respectively. Note that we have used Eckart's definition of flow velocity of the fluid.
 
 Expanding the equations in terms of derivative of temperature and chemical potential over temperature we get, 
 \ba
 \frac{\del n_{\pi}}{\del T}DT+\frac{\del n_{\pi}}{\del \left(\mu_{\pi}/T\right)}D\left(\frac{\mu_{\pi}}{T}\right)+0\, \cdot D\left(\frac{\mu_{N}}{T}\right)=-n_{\pi} \del_{\mu}u^{\mu}~~~~~~~~~~~~~~~~\nonumber\\
 \frac{\del n_{\pi}}{\del T}DT+0\cdot\,  D\left(\frac{\mu_{\pi}}{T}\right)+\frac{\del n_{N}}{\del \left(\mu_{N}/T\right)}D\left(\frac{\mu_{N}}{T}\right)=-n_{N} \del_{\mu}u^{\mu}~~~~~~~~~~~~~~~\\
 \left[n_{\pi}\frac{\del e_{\pi}}{\del T}+n_{N}\frac{\del e_{N}}{\del T}\right]DT+n_{\pi}\frac{\del e_{\pi}}{\del \left(\mu_{\pi}/T\right)}D\left(\frac{\mu_{\pi}}{T}\right)+n_{N}\frac{\del e_{N}}{\del \left(\mu_{N}/T\right)}D\left(\frac{\mu_{N}}{T}\right)=-P\del_{\mu}u^{\mu}
 \label{apC5}
 \ea
 
where $P=P_{\pi}+P_{N}$ is the total pressure. Using the relations from Appendix A we get
\ba 
\frac{\del e_{\pi}}{\del T}&=&4z_{\pi}\frac{S_{3}^1}{S_{2}^1}+
z_{\pi}\frac{S_2^2S_3^0}{(S_2^1)^2}-\frac{S_2^2}{S_2^1}+z_{\pi}^2
\left[\frac{S_2^0}{S_2^1}-\frac{S_3^1S_3^0}{(S_2^1)^2}\right]\nonumber\\
\frac{\del e_{\pi}}{\del(\mu_{\pi}/T)}&=&-T\left[1-\frac{S_2^2S_2^0}{(S_2^1)^2}\right]
+Tz_{\pi}\left[\frac{S_3^0}{S_2^1}-\frac{S_3^1S_2^0}{(S_2^1)^2}\right]~.~~~\nonumber\\
\frac{\del n_{\pi}}{\del T}&=&\frac{4\pi}{\left(2\pi\right)^{2}}T^{2}\left[-z_{\pi}^{2}S^{1}_{2}+z_{\pi}^{3}S^{0}_{3}\right]\nonumber\\
\frac{\del n_{\pi}}{\del \left(\mu_{\pi}/T\right)}&=&\frac{4\pi}{\left(2\pi\right)^{3}~}z_{\pi}^{2}T^{3}S_{2}^{0}~.
\label{apC6}
\ea
Putting in eq.~(\ref{apC5}) and solving for $DT$, $D\left(\mu_{\pi}/T\right)$ and $D\left(\mu_{N}/T\right)$ we get\footnote{To get the expression for the derivative of $e_{N}$ and $n_{N}$, $z_{\pi}$ and $S^{\alpha}_{\beta}$ are to be replaced by $z_{N}$ and $T^{\alpha}_{\beta}$ respectively.}
\ba 
T^{-1}DT&=&\left(1-\gamma^{'}\right)\del_{\mu}u^{\mu}\\
TD\left(\frac{\mu_{\pi}}{T}\right)&=&T\left[\left(\gamma_{\pi}^{''}-1\right)\hat{h_{\pi}}-\gamma_{\pi}^{'''}\right]\del_{\mu}u^{\mu}\\
TD\left(\frac{\mu_{N}}{T}\right)&=&T\left[\left(\gamma_{N}^{''}-1\right)\hat{h_{N}}-\gamma_{N}^{'''}\right]\del_{\mu}u^{\mu}
\label{apC7}
\ea 
where,
\ba 
\gamma^{'}=\frac{1}{|A|}\{g_{\pi}\left[z_{\pi}^{3}\left(4S_{2}^{0}S_{3}^{1}T_{2}^{0}+S_{2}^{1}S_{3}^{0}T_{2}^{0}\right)+z_{\pi}^{4}\left(\left(S_{2}^{0}\right)^2T_{2}^{0}-\left(S_{3}^{0}\right)^2T_{2}^{0}\right)\right]~~~~~~~~~~~~~~~\nonumber\\+g_N\left[z_{N}^{3}\left(4S_{2}^{0}T_{2}^{0}T_{3}^{1}+S_{2}^{0}T_{2}^{1}T_{3}^{0}\right)+z_{N}^{4}\left(S_{2}^{0}\left(T_{2}^{0}\right)^2-S_{2}^{0}\left(T_{3}^{0}\right)^2\right)\right]\}
\ea
 
\ba
\gamma_{\pi}^{''}=\frac{1}{|A|}\{g_{\pi}\left[-5z_{\pi}^{2}\left(S_{2}^{1}\right)^2T_{2}^{0}+z_{\pi}^{3}\left(3S_{2}^{0}S_3^1T_2^0+3S_2^1S_3^0T_2^0\right)+z_{\pi}^{4}\left(\left(S_2^0\right)^2T_2^0-\left(S_3^0\right)^2T_2^0\right)\right]\nonumber\\
+g_{N}\left[-z_{N}^{2}S_2^0\left(T_2^1\right)^2+z_{N}^3\left(3S_2^0T_2^0T_3^1+2S_2^0T_2^1T_3^0\right)
+z_{N}^4\left(S_2^0\left(T_2^0\right)^2-S_2^0\left(T_3^0\right)^2\right)\right]\}
\ea 
\ba
\gamma_{\pi}^{'''}=\frac{1}{|A|}\{g_{\pi}\left[z_{\pi}^4S_2^1S_2^0T_2^0\right]+g_N[z_{N}^3\left(4S_2^1T_2^0T_3^1+S_2^1T_2^1T_3^0\right)-z_{\pi}z_{N}^2S_{3}^{0}\left(T_2^1\right)^2~~~~~~~~~~~~~~~
\nonumber\\+z_{N}^4\left(S_2^1\left(T_2^0\right)^2-S_2^1\left(T_3^0\right)^2\right)
+z_{\pi}z_N^3\left(S_3^0T_2^1T_3^0-S_3^0T_2^0T_3^1\right)]\}
\ea

and
\ba
|A|=g_{\pi}\left[-z_{\pi}^2\left(S_2^1\right)^2T_2^0+z_{\pi}^3\left(3S_2^0S_3^1T_2^0+2S_2^1S_3^0T_2^0\right)+z_{\pi}^4\left(\left(S_2^0\right)^2T_2^0-\left(S_3^0\right)^2T_2^0\right)\right]\nonumber\\
+g_{N}\left[-z_{N}^2S_2^0\left(T_2^1\right)^2+z_{N}^3\left(3S_2^0T_2^0T_3^1+2S_2^0T_2^1T_3^0\right)+z_{N}^4\left(S_2^0\left(T_2^0\right)^2-S_2^0\left(T_3^0\right)^2\right)\right]~.
\ea
The expressions of $\gamma_{N}^{''}$ and $\gamma_{N}^{'''}$ may be obtained by replacing $S_{\beta}^{\alpha}$ with $T_{\beta}^{\alpha}$ and vice versa in $\gamma_{\pi}^{''}$ and $\gamma_{\pi}^{'''}$ respectively.

\section*{Appendix-C}
The distribution function is expanded as $f=f^{(0)}+\delta f$ around an equilibrium distribution function $f^{(0)}$ which gives the energy and particle density. We thus have 
\ba  
n=\int \frac{d^{3}p}{p^0}p^{\mu}u_{\mu}f^{(0)}~,\\
ne=\int \frac{d^{3}p}{p^0}(p^{\mu}u_{\mu})^{2}f^{(0)}~.
\ea
This leads us to the following solubility conditions that has to be satisfied by $\delta f$.
\ba  
\int \frac{d^{3}p}{p^0}p^{\mu}u_{\mu}\delta f=0~,\\
\int \frac{d^{3}p}{p^0}(p^{\mu}u_{\mu})^{2}\delta f=0~.
\ea

\end{document}